\documentstyle[11pt,newpasp,twoside,epsf]{article}
\def\edcomment#1{\iffalse\marginpar{\raggedright\sl#1\/}\else\relax\fi}

\reversemarginpar

\begin{document}
\title{The black hole mass vs bulge mass relationship in spiral galaxies}

\author{
Alessandro Marconi$\,^1\,$,
David Axon$\,^2\,$,
John Atkinson$\,^2\,$,
James Binney$\,^3\,$,
Alessandro Capetti$\,^4\,$,
Marcella Carollo$\,^5\,$,
Linda Dressel$\,^6\,$,
Holland Ford$\,^7\,$,
Joris Gerssen$\,^6\,$,
Mark Hughes$\,^2\,$,
Duccio Macchetto$\,^6\,$,
Witold Maciejewski$\,^3\,$,
Michael Merrifield$\,^8\,$,
Claudia Scarlata$\,^6\,$,
William Sparks$\,^6\,$,
Massimo Stiavelli$\,^6\,$,
Zlatan Tsvetanov$\,^7\,$,
Roeland van der Marel$\,^6\,$
}
\affil{
$^1$Osservatorio Astrofisico di Arcetri,
L.go Fermi 5, I-50125 Firenze, Italy\newline
$^2$Department of Physical Sciences, University of Hertfordshire, UK\newline
$^3$Theoretical Physics, University of Oxford, UK\newline
$^4$Osservatorio Astronomico di Torino, Italy\newline
$^5$Columbia University, Department of Astronomy, USA\newline
$^6$Space Telescope Science Institute, USA\newline
$^7$Center for Astrophysical Sciences, Johns Hopkins University, USA\newline
$^8$School of Physics \& Astronomy, University of Nottingham, UK
}

%\affil{
%$^1$Osservatorio Astrofisico di Arcetri,
%L.go Fermi 5, I-50125 Firenze, Italy\newline
%$^2$Department of Physical Sciences, University of Hertfordshire, College Lane, Hatfield, Herts AL10 9AB, UK\newline
%$^3$Theoretical Physics, 1 Keble Road, University of Oxford, Oxford, OX1 3NP, UK\newline
%$^4$Osservatorio Astronomico di Torino, Strada Osservatorio 20, I-10025 Pino Torinese, Italy\newline
%$^5$Columbia University, Department of Astronomy, New York, NY 10027, USA\newline
%$^6$Space Telescope Science Institute, 3700 San Martin Drive, Baltimore, MD 21218, USA\newline
%$^7$Center for astrophysical Sciences, Johns Hopkins University, Baltimore, MD 21218-2686, USA\newline
%$^8$School of Physics \& Astronomy, University of Nottingham, Nottingham NG7 2RD, UK
%}

\begin{abstract}
We describe an on-going HST program aimed at determining the relationship
between the nuclear black hole mass and bulge mass in spiral galaxies.  We have
selected a volume limited sample of 54 nearby spiral galaxies for which we
already have ground based emission line rotation curves, CCD surface photometry
and radio maps. We are now obtaining HST/STIS longslit observations of each of
the galaxies in the sample in order to determine the nuclear H$\alpha$ rotation
curve at high ($\sim 0\farcs1$) spatial resolution.  We will use these data to
measure the unresolved dark mass concentration at the nucleus of each object.
Here we show the first results from observations of objects in the sample.
\end{abstract}

\section{Introduction}

It has long been suspected that the most luminous AGN are powered by accretion
of matter onto massive black holes (MBH).  This
belief, combined with the observed evolution of the space-density of AGN
(van der Marel 1999 and references therein) and the high
incidence of low luminosity nuclear activity in nearby galaxies (Ho,
Filippenko, \& Sargent 1997), implies that a significant fraction of luminous
galaxies must host black holes of mass $10^6-10^{10}$M$_{\sun}$. 

It is now clear that a large fraction of hot spheroids contain a MBH (e.g.
Harms et al.\ 1994;  
Macchetto et al.\ 1997; van der Marel et al.\ 1998; Richstone et al.\ 1998).
Moreover, there is a suggestion in the data that the hole mass is proportional
to the mass (or luminosity) of the host spheroid.  Quantitatively, $M_{\rm
BH}/M_{\rm sph} \approx 0.5\%$ (e.g. Richstone et al. 1998).  This relation is still controversial, however, both
because the sensitivity of published searches is correlated with bulge
luminosity, and because there is substantial scatter in $M_{\rm BH}$ at fixed
$M_{\rm sph}$.  Recently Ferrarese \& Merrit (2000) and Gebhardt et al.\ (2000)
have shown that a tighter correlation holds between the BH mass and the
velocity dispersion of the bulge.  Clearly, any correlation of black hole and
spheroid properties would have important implications for theories of galaxy
formation in general, and bulge formation in particular.

To date the majority of BH detections have taken place in luminous, early-type
systems, and most of the data for spirals come either from studies of water
masers, or are for nearby spirals in which a MBH can be detected through stellar
dynamics. A more systematic study of spiral galaxies is clearly needed at this
stage.

\section{Description of the sample}

Prompted by the above motivations, we are undertaking a comprehensive survey
for MBHs in spiral galaxies, both quiescent and active.  We identified a volume
limited sample ($V<2000\,$km/s) of 54 Sb, SBb, Sc, and SBc spiral galaxies from
a ground-based study by Axon et al.\ who obtained H$\alpha$ and [NII] rotation
curves at seeing-limited resolution of 1\arcsec for 128
spiral galaxies from the RC3 catalogue.  The selected spiral galaxies are known
to have nuclear gas disks and span wide ranges in bulge mass and concentration.
The low redshift cut-off was chosen to ensure a high spatial resolution on source
in order to detect even low-mass MBHs. The frequency of AGN in our sample
is typical of that found in other surveys of nearby spirals, with comparable
numbers of weak nuclear radio sources and LINERS.  

\begin{figure}
\plotfiddle{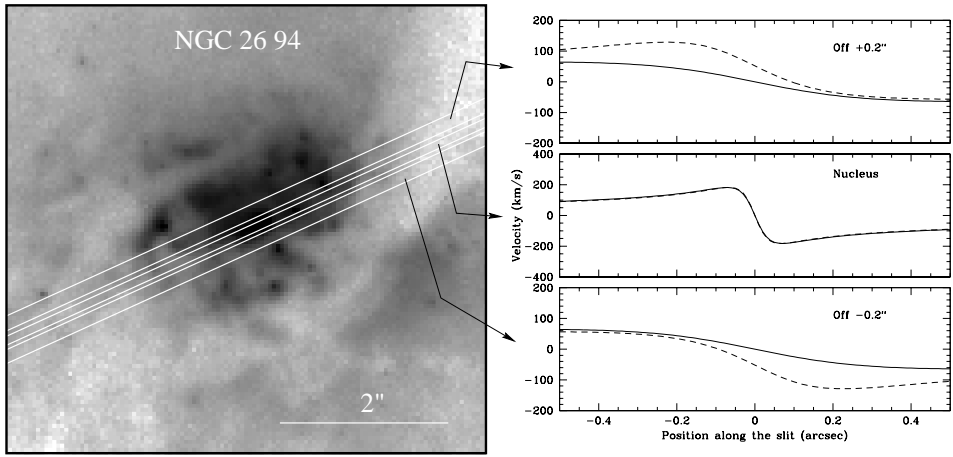}{180pt}{-90}{50}{50}{-220}{180}
\caption{Three parallel slit positions with the central one
on the galaxy nucleus constrain the line of nodes of the nuclear gas disk. The
solid and dashed lines in the right panel correspond to a combination of different BH masses and
angle between the slit and the line of nodes of the disk which yields the same rotation curve on
nucleus. This degeneracy can be resolved by looking at the off--nuclear
positions.}
\end{figure}

\section{Aims of the project}

The main aim of our survey is to verify if the correlations of the BH mass with
bulge mass and velocity dispersion (Richstone et al.\ 1998, Ferrarese \& Merrit
2000, Gebhardt et al.\ 2000) also hold for spiral galaxies.  Recently, it has
been shown that hierarchical galaxy formation models can account for these
correlations (Haehnelt \& Kauffmann 2000).  If massive spheroids form by
hierarchical merging, then the growth of MBHs may be also a hierarchical
process. Small spheroids have undergone less mergers and are thus expected to
have smaller BHs. Thus, the above correlations may hold the key to testing these
ideas about galaxy evolution.  Furthermore, one would like to compare the BH
mass with other galaxy properties such as the shape of the nuclear light
profiles; a comparison with AGN activity hosted in the galaxy would provide
constraints on the efficiency of accretion.

\section{Observations}

For each galaxy, we obtain three medium resolution spectra with STIS and the
G750M grating centered on H$\alpha$ using the 0\farcs1 or 0\farcs2 slit. The
resolution is thus ${\cal R}\sim 6000$ or 3000 respectively.  The spectra are
taken for three parallel slit positions, with the central one centered on the
galaxy nucleus and the other two offset by $\pm 0\farcs2$.  The reason for
this choice is explained in Fig.~1.
In order to remove cosmic rays
and hot pixels, 
we obtain two exposures for each slit position with the object shifted by an
integer number of detector pixels along the slit.
The use of HST guarantees a spatial resolution of about 0\farcs1 which
corresponds to 10pc on a galaxy at a distance of 20 Mpc.

\section{Preliminary results}

\begin{figure}
\plottwo{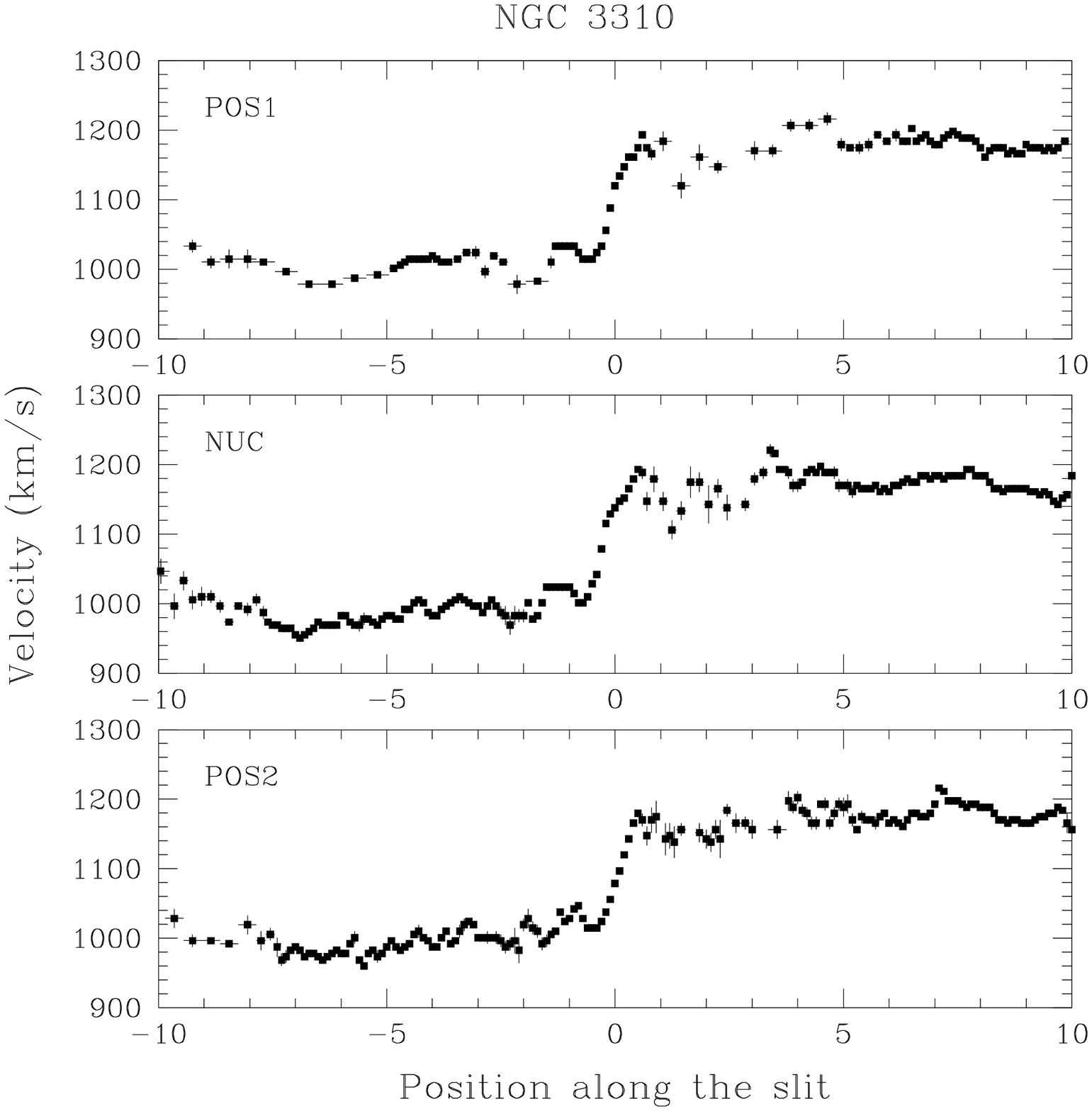}{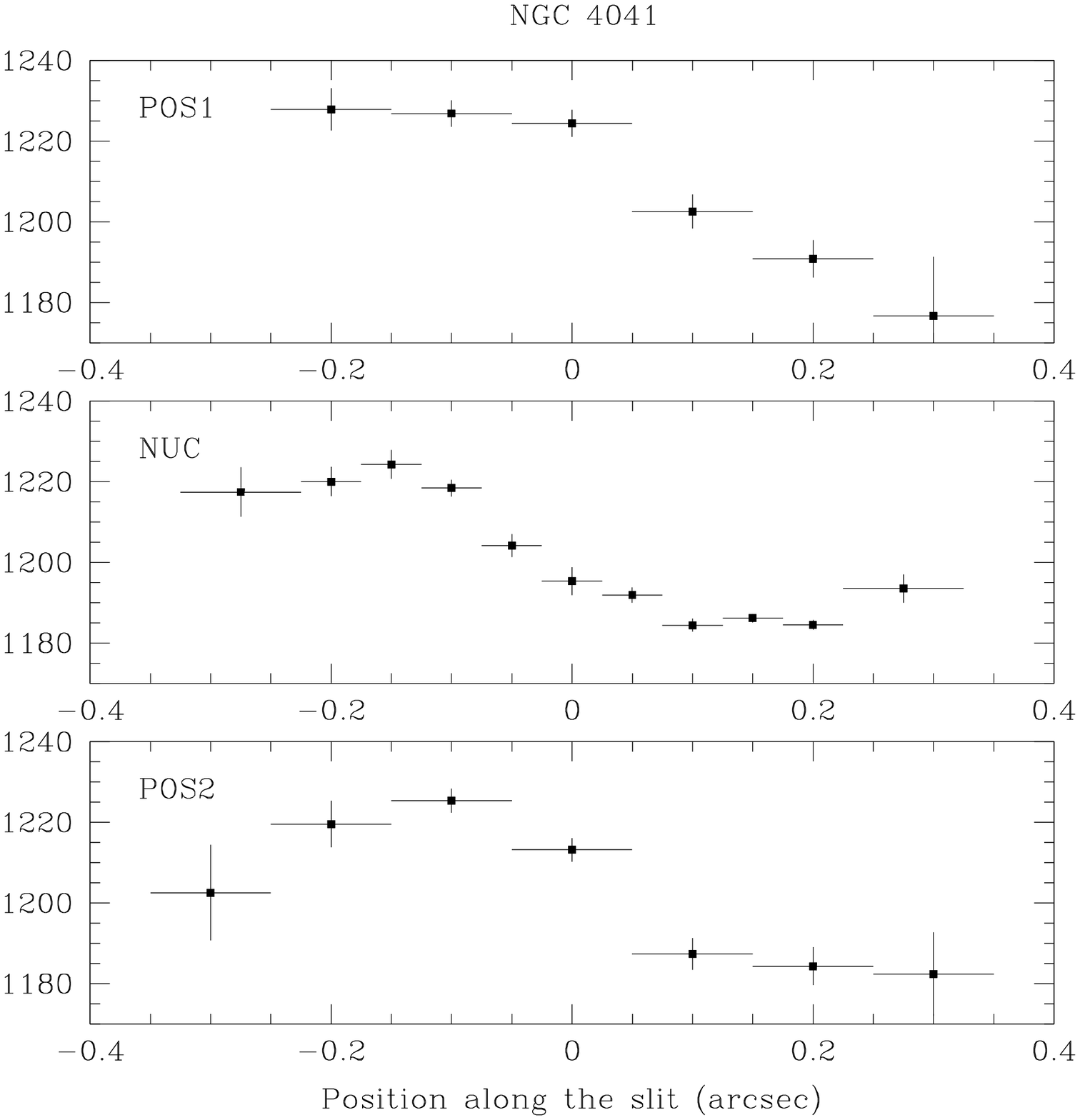}
\caption{Rotation curves of NGC 3310 (left) and NGC4041
(right). Note the small amplitude of the NGC 4041 curves ($\sim 40$
km/s).}
\end{figure}
We currently have HST/STIS data for 90\% of the sample and have extracted
rotation curves for these galaxies. Generally, the rotation curves show a steep
nuclear rise to a long plateau. In several cases the nuclear rotation curve
presents the characteristic unresolved S-shaped velocity structure expected for
Keplerian rotation around a mass concentration (Fig.~2).
This dynamical signature is accompanied by an abrupt increase
of the emission line brightness toward the center, indicative of the presence
of a well defined morphological structure, very likely a rotating circumnuclear
disk. Comparison with the stellar brightness profiles derived from the STIS
acquisition images yields stringent lower limits on the M/L ratios that
strongly suggest the presence of a supermassive black hole.
To interpret the observed rotation curves we have software in place to
calculate and fit to the data the predictions of thin disk models with
circular rotation (e.g., Marconi et al.\ 2000; van der Marel \& van
den Bosch 1998). 

As an example, we now focus on NGC 4041, which clearly shows the possibilities
of this project. 
NGC 4041 is at a distance of about 20 Mpc. The rotation curve shown in the right panel
of Fig.~2 shows the S-shaped signature of a rotating disk of
$\sim 30$ pc radius. Note the small velocity amplitude of the rotation curve
which is about 40 km/s peak to peak.
Assuming an edge-on disk with the line of nodes parallel to the slit
one can infer that the minimum dynamical mass required to explain the data
is $\sim 3\times 10^6$M$_\odot$.
This mass estimate takes into account the "smearing" of the rotation curve due
to the finite slit size and instrumental point spread function.  This low mass
value gives an indication of the lowest masses which can be detected with our
project. The inferred dynamical mass for NGC
4041 should be considered an upper limit to the mass of a possible MBH,
since the nuclear star cluster in this galaxy may be responsible for
all or part of the dynamically inferred mass.

\end{document}